\documentclass[11pt]{article}

\usepackage[utf8]{inputenc}
\usepackage{amsmath,amssymb}
\usepackage{amsfonts,amsthm}
\usepackage{bbm}
\usepackage[font=small]{caption}
\DeclareCaptionLabelFormat{previous-page}{#1 #2 \emph{(continued)}}
\usepackage{framed}

\usepackage{newfloat}
\DeclareFloatingEnvironment[fileext=lof,listname={List of Boxes}, name=Box, placement=t]{mybox}

\usepackage{array}
\usepackage[table]{xcolor} 
\definecolor{xlightgrey}{HTML}{d1dbdc}  
\definecolor{xlightblue}{HTML}{edf2fb}  
\definecolor{xlightyellow}{HTML}{fff1e6}  
\definecolor{xlightgreen}{HTML}{e9f5db}  
\definecolor{xlightbrown}{HTML}{faedcd}                              
\usepackage[framemethod=TikZ]{mdframed}
\mdfdefinestyle{box}{%
    linecolor=black,
    outerlinewidth=1pt,
    roundcorner=20pt,
    innertopmargin=\baselineskip,
    innerbottommargin=\baselineskip,
    innerrightmargin=20pt,
    innerleftmargin=20pt,
    backgroundcolor=gray!10!white}
    
\usepackage{graphicx}
\usepackage{xcolor}
\usepackage{url}
\usepackage[toc,page]{appendix} 
\usepackage{float} 
\usepackage[margin=1in]{geometry}

\usepackage{enumitem}
\usepackage[square,numbers,sort&compress]{natbib}

\usepackage[colorlinks=true,
            linkcolor=blue,
            urlcolor=blue,
            citecolor=blue]{hyperref}
\usepackage{algorithm,algorithmic}
\usepackage{booktabs}
\usepackage{tabularx}

\theoremstyle{plain}

\newcommand{\tref}[1]{Table~\ref{#1}}
\newcommand{\fref}[1]{Figure~\ref{#1}}

\def\bbP{\mathbb{P}}
\def\bbR{\mathbb{R}}

\title{Genomic Language Models: Opportunities and Challenges}
\author{Gonzalo Benegas$^{1,}$\footnote{These authors contributed equally to this work.}\;, Chengzhong Ye$^{2,*}$, Carlos Albors$^{1,*}$, Jianan Canal Li$^{1,*}$, Yun S. Song$^{1,2,3,}$\footnote{To whom correspondence should be addressed: yss@berkeley.edu}\\[5mm]
$^1$Computer Science Division, University of California, Berkeley\\
$^2$Department of Statistics, University of California, Berkeley\\
$^3$Center for Computational Biology, University of California, Berkeley}
\date{\today}

\date{\today}

\usepackage{multibib}
\newcites{method}{References}
\usepackage{nameref}

\begin{document}

\maketitle 

\begin{abstract}
Large language models (LLMs) are having transformative impacts across a wide range of scientific fields, particularly in the biomedical sciences. Just as the goal of Natural Language Processing is to understand sequences of words, a major objective in biology is to understand biological sequences. Genomic Language Models (gLMs), which are LLMs trained on DNA sequences, have the potential to significantly advance our understanding of genomes and how DNA elements at various scales interact to give rise to complex functions. To showcase this potential, we highlight key applications of gLMs, including functional constraint prediction, sequence design, and transfer learning. Despite notable recent progress, however, developing effective and efficient gLMs presents numerous challenges, especially for species with large, complex genomes. Here, we discuss major considerations for developing and evaluating gLMs.
\end{abstract}

\clearpage
\section*{INTRODUCTION}
Recent advances in AI/ML have profoundly impacted a wide range of scientific disciplines, revolutionizing approaches to modeling, data analysis, interpretation, and discovery.  One of the key pillars of this development is self-supervised learning, in which training on massive amounts of unlabeled data enables the learning of complex features and their interactions.  This paradigm has particularly transformed Natural Language Processing (NLP), allowing AI models to match human performance on several challenging tasks, including translation \cite{transformer}, speech recognition \cite{gulati2020conformer}, and even answering questions from standardized professional and academic exams~\cite{openai2024gpt4technicalreport}.

Just as the aim of NLP is to understand sequences of natural language, a major aim of computational biology is to understand biological sequences. As such, there has been intense recent interest in adapting modern techniques from NLP for biological sequences (DNA, RNA, proteins).  In particular, protein sequence databases (e.g., UniProt \cite{uniprot}) have grown exponentially over the past decade, and protein language models (pLMs) trained on these immense data have achieved impressive performance on complex problems such as structure prediction \cite{lin2023evolutionary} and variant effect prediction \cite{esm1v,truong2023poet}, to name just a few examples (see \cite{bepler2021learning,ruffolo2024designing} for reviews on pLMs and their applications).
This success aligns with the intuition that billions of years of evolution have explored portions of the protein sequence space that are relevant to life, so large unlabeled datasets of protein sequences are expected to contain significant biological information. 

In a similar vein, large language models (LLMs) trained on DNA sequences have the potential to transform genomics, but training an effective model for genomes presents several additional challenges.  For instance, unlike proteins, which are functionally important units and relatively small in size, most genomes are much larger and often contain vast amounts of complex, non-functional regions that overshadow the amount of functional elements.  In addition, the number of available whole-genome sequences across the tree of life is minuscule compared to the hundreds of millions of protein sequences, limiting the diversity of functionally important DNA elements in training data.
Despite these issues, we believe that language models trained on genomes -- referred to as genomic language models (gLMs) -- hold great promise for biology.  In this article, we review some of the key opportunities and challenges in this domain, and outline major considerations that should be addressed to develop and evaluate gLMs that would be useful to the genomics community.

\begin{mybox}[p]
\begin{mdframed}[style=box]
\caption{General Language Model Framework.}
\label{box:LM_framework}
At a high level, a language model is trained to learn the conditional probability distribution of the form $\bbP[X_i \mid X_{-\text{Masked}}]$ for $i\in \text{Masked}$ (in \textbf{Masked Language Modeling, MLM)} or  $\bbP[X_k\mid X_{1:k-1}]$ (in \textbf{Causal Language Modeling, CLM}), where $X = (X_1,X_2,\ldots)$ denotes a sequence of ``tokens'' (e.g., nucleotides or amino acids) and ``Masked'' denotes a collection of masked positions.  
The key to recent advances in NLP is that, instead of fitting a simple parametric model of context dependency that one designs by hand, one lets the data speak for themselves and fit more complex models as more data are observed, by leveraging powerful deep learning architectures.  
\fref{fig:overview} depicts the language modeling framework for DNA.
While the model is trained to predict the nucleotide at each masked site using information from unmasked sites, it will learn position-specific contextual representation (called embedding, a high-dimensional vector in $\bbR^n$), which then gets converted into a probability distribution over \texttt{\{A,C,G,T\}}.  These embeddings and probability distributions, both of which are position-specific, can be applied to many problems in genomics.

\begin{figure}[H]
    \centering
    \includegraphics[width=\textwidth]{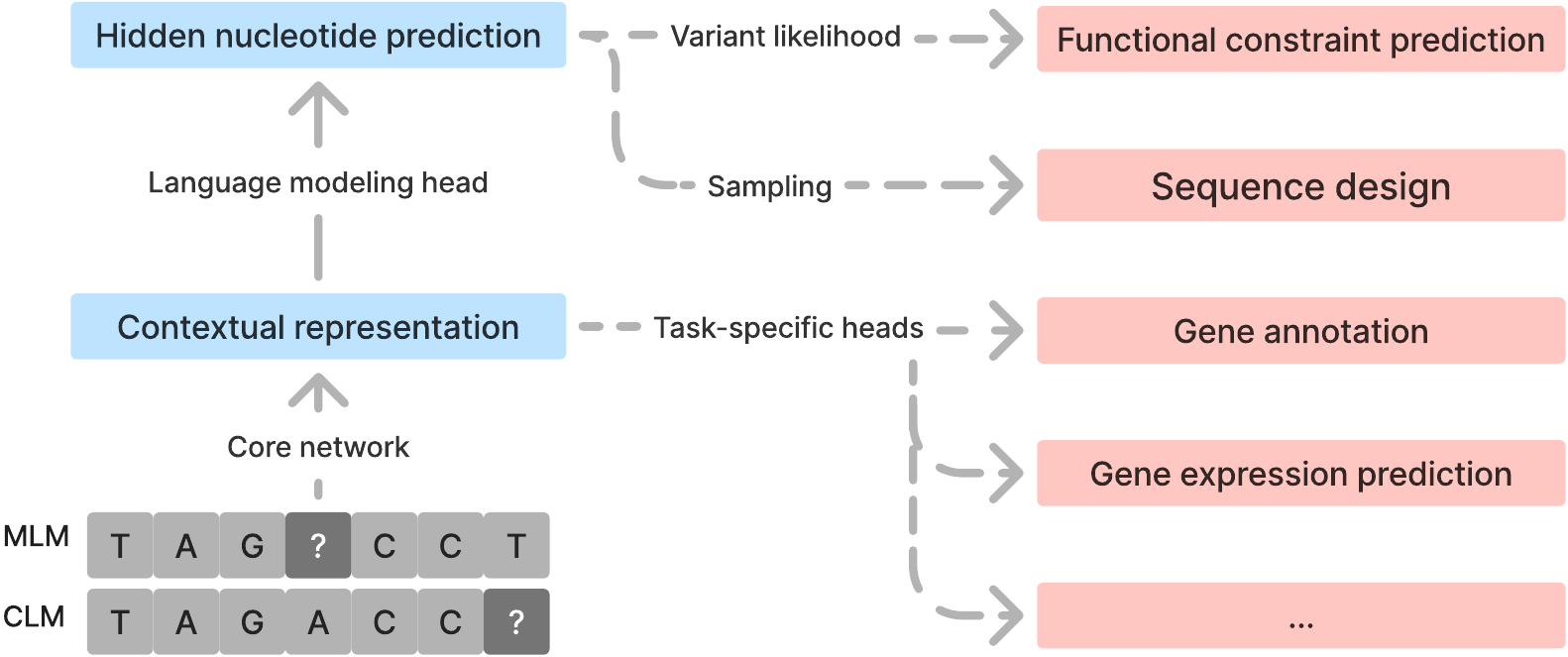}
\caption{
\textbf{Training and applications of gLMs.}  The schematic on the left-hand side illustrates gLM training.
The log-likelihood ratio (LLR) between two alleles (specifically,\break $\log[\bbP(X_i = a \mid X_{-i})/\bbP(X_i = b \mid X_{-i})]$) is a good unsupervised predictor of functional constraint (\nameref{sec:functional-constraint}).
New sequences can be generated by sampling from the learned probability distribution (\nameref{sec:generation}).
A vector representation, called embedding, of each token in the input sequence can be extracted and adapted for different downstream tasks (\nameref{sec:transfer-learning}).
}
\label{fig:overview}
\end{figure}
\end{mdframed}
\end{mybox}

\section*{APPLICATIONS}
\label{sec:applications}

\begin{figure}[h]
    \centering
    \includegraphics[width=\textwidth]{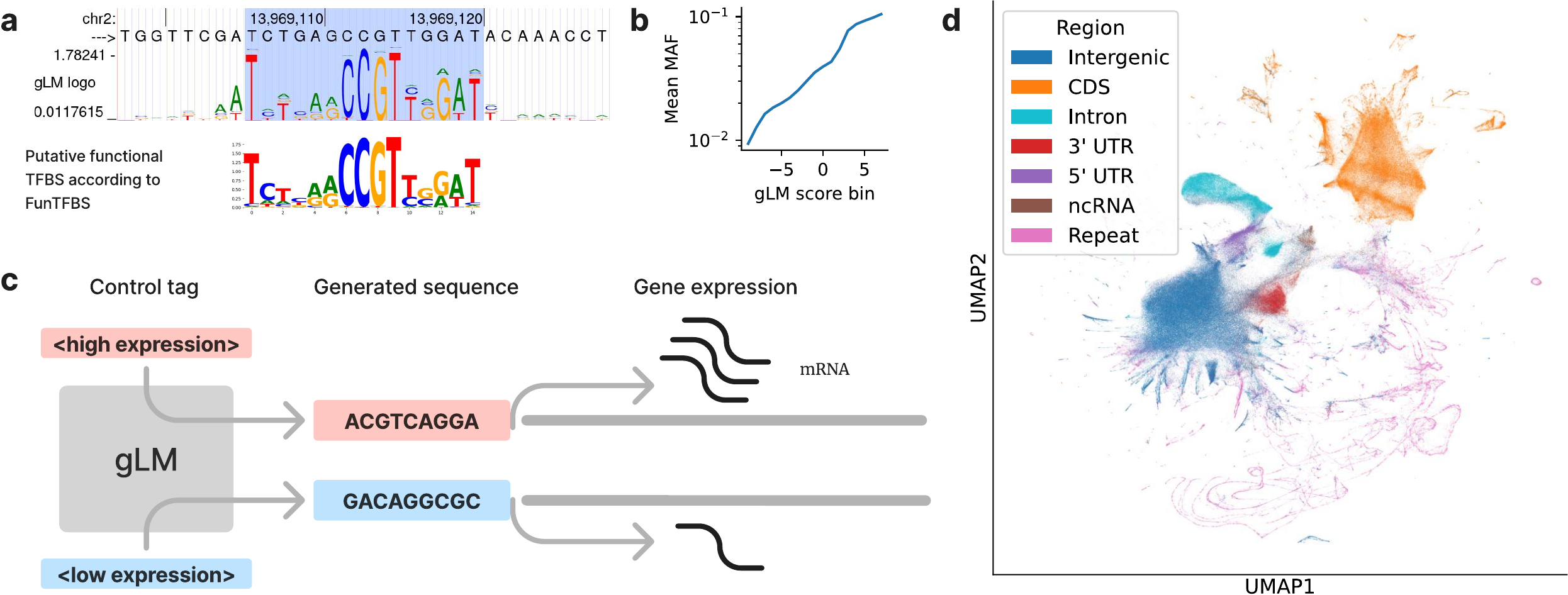}
\caption{
\textbf{Application examples.}
\textbf{(a)} gLM predicted logo plot (top) at a promoter, highlighting a motif (bottom logo) that matches a putative functional TFBS.
\textbf{(b)} Correlation between variant minor allele frequency (MAF) and gLM score (log-likelihood ratio).
\textbf{(c)} A gLM can be prompted with different control tags to design promoter sequences driving high or low expression in a given cell type.
\textbf{(d)} Visualization of gLM embeddings for different classes of genomic windows, illustrating that the learned representations contain useful information such as gene regions. 
Note: Panels a,b,d were generated using the GPN model.
}
\label{fig:applications}
\end{figure}

The general language model framework is summarized in Box~\ref{box:LM_framework}.  Below, we elaborate on three main application areas of gLMs:  \nameref{sec:functional-constraint}, \nameref{sec:generation}, and \nameref{sec:transfer-learning}.

\subsection*{Functional constraint prediction}
\label{sec:functional-constraint}

An intriguing application of gLMs is the prediction of functional constraint on a genomic locus without any supervision on the task.
A significant benefit of this approach is its independence from labels, such as whether a variant is disease-causing, which are often limited and subject to biases.
The underlying idea is that reference genomes, typically derived from healthy individuals, are relatively depleted of deleterious variants.
Consequently, models trained on these data are predisposed to assigning lower probabilities to harmful variants.
This observation underpins the strategy of using the log-likelihood ratio (LLR) between two alleles (i.e.,  $\log[\bbP(X_i = a \mid X_{-i})/\bbP(X_i = b \mid X_{-i})])$) to estimate their relative fitness.

Functional constraint prediction using the LLR was initially introduced in the context of protein sequence models, leading to outstanding results in predicting the effects of missense variants \cite{riesselman2018deep,eve,esm1v,brandes2023}.
Expanding this approach, genome-wide functional constraint prediction using a gLM was first undertaken by GPN~\cite{gpn}, achieving state-of-the-art results in the model plant \textit{Arabidopsis thaliana}.
To illustrate how a gLM might be able to predict functional constraint, we note that gLMs can learn transcription factor binding site (TFBS) motifs, understanding which positions are under constraint and which are not (\fref{fig:applications}a).  In addition, GPN's LLR score is correlated with allele frequencies in natural \textit{Arabidopsis thaliana} populations, even though the model was only trained on a single genome from this species (\fref{fig:applications}b).
Subsequently, AgroNT \cite{AgroNT} and PlantCaduceus \cite{PlantCaduceus} have also obtained excellent results in other plant species. 
For the human genome, however, the LLR from the Nucleotide Transformer (NT) \cite{dalla2023nucleotide} fell short of existing baselines.
Meanwhile, GPN-MSA \cite{gpnmsa}, leveraging a whole-genome multiple sequence alignment (MSA) across diverse vertebrate species, was able to attain state-of-the-art performance (see \nameref{sec:learning} for further MSA considerations).
It should be noted that the observed nucleotide distribution is driven not only by functional constraint but also by mutational biases; explicitly incorporating this information into functional constraint prediction is a promising avenue of future research.

For a single nucleotide polymorphism (SNP), the LLR can be computed in a single query to an MLM with the variant position masked, but in two queries to a CLM on the reference and alternate sequences.
A CLM can just as easily handle multiple substitutions, insertions and deletions, while an MLM must resort to a more expensive pseudo-LLR \cite{hsu2022learning2,brandes2023}.
Scores other than the LLR have been proposed for functional constraint prediction, such as the distance in embedding space \cite{dalla2023nucleotide,AgroNT} or the change in nucleotide probabilities in positions around a mutation \cite{tomaz2024nucleotide}.
While the LLR is widely used in both the pLM and gLM communities, it is important to understand better in which contexts these alternative scores are useful.

There are two main kinds of variant effect predictors in genomics: \textit{functional constraint predictors}, including gLMs and traditional conservation scores \cite{phastcons,phylop}, and \textit{activity predictors}, such as the gene expression predictor Enformer \cite{enformer} or the splicing predictor SpliceAI \cite{SpliceAI}.
These two kinds of models are related in the sense that if a variant at a locus is under selection, it induces a change in activity in some context (e.g., change in transcription of a certain gene during limb development), ultimately affecting a high-level trait (e.g., polydactyly).
Functional constraint models cover all possible mechanisms and contexts that affect the overall organismal fitness, while activity models reflect only those they are explicitly trained on (some data, such as protein expression in the developing human brain, are just hard to obtain).
On the other hand, activity models can nominate a specific mechanism and context through which a variant acts, while functional constraint models do not offer a mechanistic interpretation.

With regards to functional variant prioritization, there are some additional considerations.
An activity model typically gives similar scores to two variants that induce a similar expression fold-change but in different genes, even if there is a vast difference in physiological tolerance to their expression levels.
On the other hand, a trait not under detectable selection could still be of scientific or medical interest.
In this case, a functional constraint model would have limited power to prioritize variants affecting it, especially if they have small effect sizes, as is the case in complex trait GWAS.
However, while a gLM's LLR might not have high power in this setting, gLM's learned embeddings (Box~\ref{box:LM_framework}) could still provide value with additional supervision on labeled data \cite{caduceus}.

\subsection*{Sequence design}
\label{sec:generation}

Designing novel biological sequences is of great interest to both the academic and industry research communities due to its immense potential in drug discovery and delivery; agricultural improvement; bioremediation; and the development of biological research tools.
We here describe sequence generation with a CLM (Box~\ref{box:LM_framework}) as it is the most common approach (see \nameref{sec:learning} for generation with MLMs).
Specifically, the sequence generation task is decomposed into a series of next-token prediction problems.
Starting with a given sequence fragment (referred to as prompts~\cite{brown2020language}, or control tags \cite{madani2023large}), the language model can predict the next token recursively and generate a whole new sequence.
pLMs have been shown to be powerful tools for protein design~\cite{madani2023large, ingraham2019generative, hsu2022learning, shin2021protein}. Going beyond coding sequences, designing non-coding sequences is also crucial due to its applications such as gene and cell therapies \cite{lal2024reglm}, as well as synthetic biology \cite{evo}.  Such design tasks have previously been addressed using supervised activity models \cite{wang2020synthetic, jores2021synthetic, de2022deepstarr}, but more recently several works have explored the use of gLMs to tackle this challenge as described below.

The model regLM \cite{lal2024reglm} was built upon the causal gLM HyenaDNA \cite{hyenadna} and used to perform \textit{de novo} generation of promoter and enhancer sequences. HyenaDNA models are trained or fine-tuned on regulatory sequences with control tags prepended. The trained model is then used to generate new regulatory sequences with given tags (\fref{fig:applications}c). The authors performed \textit{in silico} evaluation of the diversity and activity of the generated sequences in yeast and human cell lines, and demonstrated the sequences to have desired functionality as well as realistic and diverse sequence features.

gLMs have the unique potential for multi-modal design tasks such as generating protein-RNA complexes by unifying them as DNA sequence design. For instance, EVO, a gLM trained on prokaryote genomes, was used to design novel CRISPR-Cas systems \cite{evo}. The model was fine-tuned using a dataset of CRISPR-Cas sequences with Cas subtype-specific prompt prepended. The fine-tuned model was able to generate novel CRISPR-Cas sequences that matched the subtype prompt and had predicted structures that resemble naturally existing systems.

Additionally, gLMs can be potentially used to design organized, functional DNA sequences at the chromosome or genome-scale. Recently, two gLMs, MegaDNA and EVO, have explored such design tasks for prokaryote genomes \cite{megadna, evo}. EVO was used to generate 20 sequences of size about 650~Mbp. The generated sequences were found to have realistic coding sequence density, protein sequences with predicted secondary structure and globular folds, as well as plausible tRNA sequences. MegaDNA was used to generate full bacteriophage genomes up to 96~kbp. Apart from validating coding sequences, the author further identified functional regulatory elements including promoters and ribosome binding sites in the generated sequences. Yet, such mega-scale DNA sequence design tasks remain challenging. The generated sequences by EVO were found to lack highly conserved marker genes that typically exist in functional prokaryote genomes, and the predicted protein structures have limited matches to natural protein databases. A recent independent evaluation \cite{ratcliff2024transformer} revealed that the sequence composition of MegaDNA-generated genomes is still largely dissimilar to natural genomes. Therefore, further work is needed to refine the methods to enable \textit{de novo} design of fully functional genomes with gLMs.

\subsection*{Transfer learning}
\label{sec:transfer-learning}

Neural networks trained to predict annotations from functional genomics experiments have been widely utilized to interpret the functions of genomic elements. A significant application has been predicting variant effects on molecular phenotypes, such as gene expression \cite{deepbind, deepsea, basset, Basenji, enformer} and splicing \cite{SpliceAI, zeng2022pangolin}. The ability of neural networks to interpret complex interactions between genomic sites has made them essential tools for tackling these important problems, but suitable training data are often difficult to collect and consequently limited.  To generalize on prediction tasks, models need to be capable of identifying the broad set of functionally important sequence elements, which may require substantial data and computation. To overcome the limitations of insufficient data for individual tasks, developers have employed \textit{transfer learning} methods --- techniques that leverage knowledge gained from training models on one task to improve performance on related tasks. Specifically, most neural networks trained to predict functional annotations have been trained to predict a wide array of annotations simultaneously, forcing these models to learn a single unifying representation. This, in turn, has improved their generalization performance.

\begin{mybox}[t]
\begin{mdframed}[style=box]
\caption{Transfer Learning in NLP}
\label{box:transfer-learning-NLP}
For NLP models to generalize on most tasks (including typical tasks, such as sentiment analysis, question answering, and part-of-speech tagging, to name only a few), models need to understand grammar and meaning. However, data specific to these tasks are limited. Utilizing LLMs trained on raw text data (sourced from articles, books, and websites) for transfer learning has enabled breakthrough progress on these problems \cite{devlin2018bert}. Today, virtually every state-of-the-art NLP model is adapted from an LLM.

\medskip
Transfer learning techniques have underpinned the recent boom in natural language models. In particular, the availability of pretrained models that are broadly adaptable to downstream tasks---termed ``foundation models''---has yielded a major shift in how machine learning models are developed \cite{bommasani2021a}.

\end{mdframed}
\end{mybox}

Language models may also be utilized for transfer learning. 
(See Box~\ref{box:transfer-learning-NLP} for a discussion of the utility of transfer learning for NLP.)
One technique is \textit{feature extraction}:
while learning to predict the context-dependent distribution of nucleotides, gLMs 
transform input genomic sequences into intermediate vector representations 
(Box~\ref{box:LM_framework}). 
These representations may distill relevant information, and, therefore, be utilized as features for another model.
For example, visualization of gLM embeddings reveals that, without any supervision, the model has learned to distinguish different classes of genomic elements such as coding sequence and untranslated regions \cite{gpn} (\fref{fig:applications}d). 
Embeddings from different layers can provide information useful for different tasks \cite{west2024diverse}.
Another way to utilize language models for transfer learning is to use them as \textit{pretrained} models: that is, to continue training them on downstream tasks. This technique is called \textit{fine-tuning}. Fine-tuning a pretrained neural network on a task implicitly regularizes its parameters such that the network's predictions synthesize knowledge from both tasks. As a result, pretraining neural networks tends to improve their generalization performance on downstream tasks. In recent work, SegmentNT, a model developed by fine-tuning the Nucleotide Transformer (NT) gLM \cite{dalla2023nucleotide} to the task of annotating genes and cis-regulatory elements, achieved state-of-the-art performance on this task \cite{de2024segmentnt}. Utilizing a pretrained model was shown to be critical to its success. Similarly, AgroNT \cite{AgroNT}, another model of the NT family, was pretrained on diverse plant species and then fine-tuned to predict chromatin accessibility and gene expression on select crop species. DNABERT-S \cite{dnabert-s} applied contrastive learning with pretrained DNABERT-2 \cite{dnabert2} embeddings to perform metagenomics binning.
IsoFormer \cite{garau2024multi} is an example of multi-modal transfer learning between DNA and protein language models for the task of predicting transcript isoform expression.
These recent successes suggest that fine-tuned gLMs may make meaningful progress on diverse genome interpretation tasks.

Two recent studies evaluated several gLMs in prediction tasks in the human genome and found that they generally did not outperform non-gLM baselines \cite{marin2024bend,tang2024evaluating}. These results were based on frozen embeddings; evaluating full fine-tuning would provide additional insights. While gLMs are already well suited to demonstrate the value of transfer learning in less-studied organisms, further innovation may be required for them to offer significant value in human genetics, where high-quality labeled data and carefully crafted models already exist. An important question is how far the scaling hypothesis holds for gLMs, i.e., how much increasing unlabeled data and computation will keep improving model performance. 
A recent pLM study found that scaling improved only protein structure prediction but not most other tasks such as function or property prediction \cite{li2024feature}, so gLM tasks should also be held to the same scrutiny.

\section*{DEVELOPMENT}

\begin{figure}[b!]
    \centering
    \includegraphics[width=\textwidth]{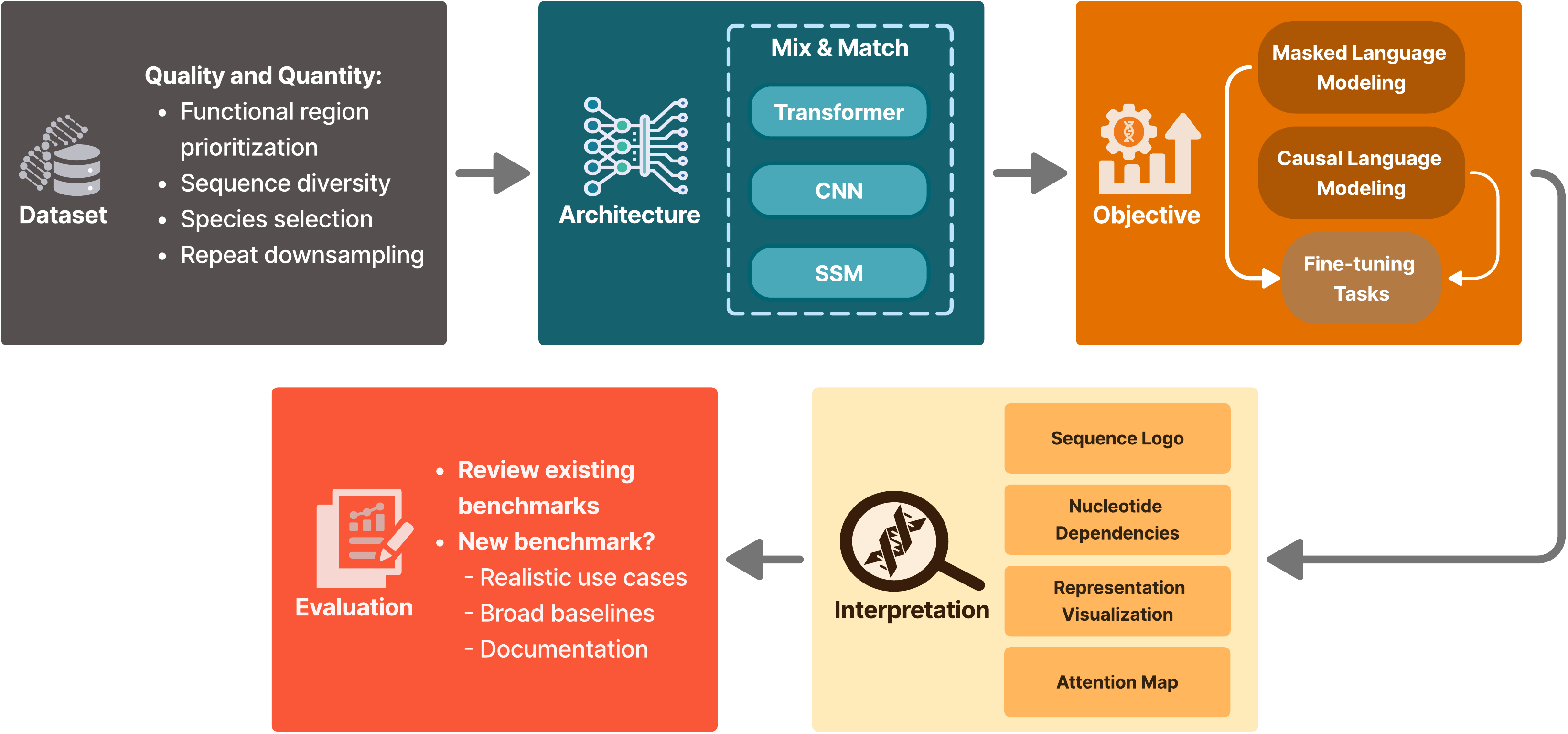}
\caption{
\textbf{Development Pipeline.}
This figure illustrates the general gLM development pipeline described in this review, from model conception to deployment. We begin with the selection and preparation of the training dataset, emphasizing the importance of data quality and quantity (\nameref{sec:learning-data}). Subsequently, in \nameref{sec:architecture} and \nameref{sec:learning}, we explore the various choices for designing and  training gLMs, discussing the strengths and weaknesses of different approaches. We also examine how hybrid models combine elements from multiple architectures to mitigate specific limitations.  In \nameref{sec:Interpretation}, we discuss methods for analyzing and interpreting the outputs of gLMs. Finally, in \nameref{sec:evaluation}, we present evaluation methods through current benchmarks, emphasizing the complexities in aligning model performance with actual biological functions.
}

\label{fig:develop}
\end{figure}

We now describe the key components of developing useful gLMs; a schematic diagram summarizing the development pipeline is illustrated in \fref{fig:develop}.  We first describe the importance of selecting and preparing training data, and then discuss architectural and training decisions.  We then consider interpreting and benchmarking gLMs.  Our aim is to provide insights into the methodologies and challenges encountered in developing gLMs that are both effective and efficient.  To provide a comprehensive view of the current landscape in the field, we list in \tref{tab:existing_gLMs} some of the existing gLMs that we are aware of and summarize their design decisions.

\begin{table}[!phbt]
    \centering
\scalebox{0.56}{\begin{tabular}{|>{\vspace{1pt}}m{4.6cm}<{\vspace{1pt}}|>{\vspace{1pt}}m{6.0cm}<{\vspace{1pt}}|>{\vspace{1pt}}m{1.2cm}<{\vspace{1pt}}|>{\vspace{1pt}}m{3cm}<{\vspace{1pt}}|>{\vspace{1pt}}m{4cm}<{\vspace{1pt}}|>{\vspace{1pt}}m{8cm}<{\vspace{1pt}}|}

\hline 
\rowcolor{xlightgrey}
 \vspace{4pt} \textbf{Model Name} \vspace{4pt} &\textbf{Pretraining data sources} &\textbf{Task} &\textbf{Architecture} &\textbf{Tokenization} &\textbf{Notes}  \\ \hline
\rowcolor{xlightblue}
\textbf{BigBird} \cite{bigbird} &Human &MLM &Transformer &BPE &~ \\ \hline
\rowcolor{xlightblue}
\textbf{DNABERT} \cite{dnabert} &Human &MLM &Transformer &overlapping $k$-mer &~ \\ \hline
\rowcolor{xlightblue}
\textbf{GeneBERT} \cite{GeneBERT} &Human &MLM &Transformer &overlapping $k$-mer &Trained to also predict chromatin accessibility ATAC-seq data. \\ \hline
\rowcolor{xlightblue}
\textbf{Epigenomic BERT} \cite{trotter2021epigenomic} &Human &MLM &Transformer &non-overlapping $k$-mer &DNA sequences are paired with associated epigenetic state information (IDEAS) \cite{zhang2016jointly} during training. \\ \hline
\rowcolor{xlightblue}
\textbf{LookingGlass} \cite{LookingGlass} &Bacteria + archaea &CLM &Recurrent Neural Network &nucleotide-level &Metagenomic sequences from diverse environments rather than assembled genomes are used for training. \\ \hline
\rowcolor{xlightblue}
\textbf{LOGO} \cite{LOGO} &Human &MLM &CNN + Transformer &overlapping $k$-mer &~ \\ \hline
\rowcolor{xlightblue}
\textbf{ViBE} \cite{gwak2022vibe} &Virus &MLM &Transformer &overlapping $k$-mer &~ \\ \hline
\rowcolor{xlightblue}
\textbf{GPN} \cite{gpn} &\textit{Arabidopsis thaliana} + 7 related Brassicales genomes &MLM &CNN &nucleotide-level &~ \\ \hline
\rowcolor{xlightblue}
\textbf{FloraBERT} \cite{levy2022florabert} &Several hundred plants + selected maize genomes &MLM &Transformer &BPE &Only 1kb promoter sequences are used in training. \\ \hline
\rowcolor{xlightblue}
\textbf{INHERIT} \cite{bai2022identification} &Bacteria + bacteriophage &MLM &Transformer &overlapping $k$-mer &~ \\ \hline
\rowcolor{xlightblue}
\textbf{GenSLMs} \cite{GenSLMs} &Prokaryotic gene sequences + SARS-CoV-2 genomes &CLM &Transformer &non-overlapping $k$-mer &Pretrain on prokaryotic genes and fine-tune on SARS-CoV-2 genomes. \\ \hline
\rowcolor{xlightblue}
\textbf{NT} \cite{dalla2023nucleotide} &Human + 1000 Genomes Project + multi-species &MLM &Transformer &non-overlapping $k$-mer &~ \\ \hline
\rowcolor{xlightblue}
\textbf{SpliceBERT} \cite{SpliceBERT} &Human + 71 vertebrate genomes &MLM &Transformer &nucleotide-level &Only RNA Transcripts are used in training. \\ \hline
\rowcolor{xlightblue}
\textbf{SpeciesLM Fungi} \cite{speciesLM} &1500 fungal genomes &MLM &Transformer &overlapping $k$-mer &Only $5'$ and $3'$ UTR regions are used in training: the $5'$ species LM and $3'$ species LM. \\ \hline
\rowcolor{xlightblue}
\textbf{GENA-LM} \cite{GENA-LM} &Human + multi-species &MLM &Transformer &BPE &~ \\ \hline
\rowcolor{xlightblue}
\textbf{DNABERT-2} \cite{dnabert2} &Human + multi-species &MLM &Transformer &BPE &~ \\ \hline
\rowcolor{xlightblue}
\textbf{HyenaDNA} \cite{hyenadna} &Human &CLM &SSM &nucleotide-level &~ \\ \hline
\rowcolor{xlightblue}
\textbf{GROVER} \cite{sanabria2023human} &Human &MLM &Transformer &BPE &~ \\ \hline
\rowcolor{xlightblue}
\textbf{DNAGPT} \cite{dnagpt} &Human + multi-species &CLM &Transformer &non-overlapping $k$-mer &~ \\ \hline
\rowcolor{xlightblue}
\textbf{GPN-MSA} \cite{gpnmsa} &Human + Multiple Sequence Alignment (MSA) with 100 vertebrate genomes &MLM &Transformer &nucleotide-level &~ \\ \hline
\rowcolor{xlightblue}
\textbf{UTR-LM} \cite{UTR-LM} &Human + 4 vertebrate genomes &MLM &Transformer &nucleotide-level &Only $5'$ UTR regions are used in training. Trained also to predict mRNA minimum free energy and secondary structures calculated by ViennaRNA \cite{viennarna}. \\ \hline
\rowcolor{xlightblue}
\textbf{hgT5} \cite{robson2023guanine} &Human &T5 \cite{raffel2020exploring} &Transformer &Unigram model \cite{kudo2018subword} &~ \\ \hline
\rowcolor{xlightblue}
\textbf{AgroNT} \cite{AgroNT} &48 plant genomes focusing on edible plant species &MLM &Transformer &non-overlapping $k$-mer &~ \\ \hline
\rowcolor{xlightblue}
\textbf{MegaDNA} \cite{megadna} &$\sim$100k bacteriophage genomes &CLM &Transformer &nucleotide-level &~ \\ \hline
\rowcolor{xlightblue}
\textbf{regLM} \cite{lal2024reglm} &Human + yeast &CLM &SSM  &nucleotide-level &Human enhancer and yeast promoter sequences are used to fine-tune/pretrain separate HyenaDNA \cite{hyenadna} models \\ \hline
\rowcolor{xlightblue}
\textbf{EVO} \cite{evo} &Bacteria + archaea + virus + plasmid &CLM &SSM + Transformer &nucleotide-level &~ \\ \hline
\rowcolor{xlightblue}
\textbf{Caduceus} \cite{caduceus} &Human &MLM &SSM &nucleotide-level &~ \\ \hline
\rowcolor{xlightblue}
\textbf{ChatNT} \cite{chatNT} &Genomic sequences + English instructions &CLM &Transformer &overlapping $k$-mer &Combines the pretrained gLM NT \cite{dalla2023nucleotide} and the English LM Vicuna \cite{vicuna}. Trained to perform all supervised genomics prediction tasks as text-to-text tasks. \\ \hline
\rowcolor{xlightblue}
\textbf{LucaOne} \cite{he2024lucaone} &Genomic and protein sequences from 169,861 species &MLM &Transformer &nucleotide- and amino acid-level &Mixed pretraining with DNA, RNA, and protein sequences. Trained also to predict 8 types of selected annotations. \\ \hline
\rowcolor{xlightblue}
\textbf{PlantCaduceus} \cite{PlantCaduceus} &16  Angiosperm genomes &MLM &SSM &nucleotide-level &~ \\ \hline
\rowcolor{xlightblue}
\textbf{CD-GPT} \cite{zhu2024cd} &Genomic and protein sequences of 14 organisms &CLM &Transformer &BPE &Mixed pretraining with DNA, RNA, and protein sequences, followed by targeted DNA-Protein and mRNA-Protein paired pretraining. \\ \hline
\rowcolor{xlightblue}
\textbf{SpeciesLM Metazoa} \cite{tomaz2024nucleotide} &494 metazoan genomes &MLM &Transformer &overlapping $k$-mer &Only trained on 2~kb upstream of start codons \\ \hline
\rowcolor{xlightblue}
\textbf{gLM2} \cite{cornman2024omg} &Metagenomes and genomes from IMG \cite{markowitz2012img} and MGnify \cite{richardson2023mgnify}.  &MLM &Transformer &BPE for nucleotides, amino acid-level for proteins &Pretraining with a mixed-modality dataset, comprising interleaved protein-coding (amino acid) and intergenic (nucleotide) sequences.  \\ \hline

\end{tabular}}
\caption{\textbf{A summary of existing gLMs.} An overview of various gLMs is provided, highlighting their pretraining datasets, tasks, architectures, tokenization methods, and unique features. The models are listed in the order of their public release dates. Abbreviations used include SSM for State Space Model, CNN for Convolutional Neural Network, BPE for Byte-Pair Encoding, CLM for Causal Language Modeling, and MLM for Masked Language Modeling.}

\label{tab:existing_gLMs}
    
\end{table}


\subsection*{Training data}
\label{sec:learning-data}

The performance of a machine learning model is significantly influenced by both its architecture and its training data.
Various model architectures such as convolutional neural networks (CNNs), Transformers, and state-space models have been successfully adapted to a wide range of domains, including natural language, images, audio, proteins, and genomics.
However, selecting suitable data for pretraining requires a deep understanding of the specific domain, especially in genomics where there is no universally accepted, curated dataset comparable to those in NLP (e.g., the Pile \cite{pile,fmcheatsheet}) or protein biology (e.g., UniProt \cite{uniprot}).

A key consideration is data quality.
For example, in NLP this may refer to data sources that have undergone editing or peer review, such as scientific articles or books \cite{fmcheatsheet}.
In the case of proteins, quality control involves removing predicted pseudogenes or truncated proteins that are no longer functional \cite{uniprot}.
However, a recent study found only 3.3\% of the bases in the human reference genome, the most popular gLM training dataset (\tref{tab:existing_gLMs}), to be significantly constrained and likely functional \cite{sullivan2023leveraging}.
Importantly, a typical genomic sequence used for training a gLM will contain a mix of functional and non-functional sites, and one cannot always separate training examples into high vs. low quality.
A proposed solution is to have a base-pair-level weighting of the training loss according to the evidence for functionality \cite{gpnmsa}.

It is standard in NLP and proteins to filter out duplicated sequences, which improves training efficiency and reduces memorization \cite{lee2021deduplicating}.
Despite the fact that a staggering 50\% of the human genome is repetitive (a high proportion across eukaryotes), very few gLM studies propose a solution (downweighting \cite{gpn, PlantCaduceus} or downsampling \cite{SpliceBERT,robson2023guanine}), let alone acknowledge the issue.
It would be insightful if studies of language model perplexity \cite{hyenadna,caduceus} would also report it separately for non-repetitive regions, to distinguish improvements due to generalization vs. memorization.

Another key question is how to ensure that the amount of data is enough.
It is likely that a single genome might not be enough to train a large model, especially if non-functional regions are downsampled or downweighted.
One approach is to add sequence variants from the same species \cite{dalla2023nucleotide}.
However, in many species, including humans, there is relatively little variation between individuals.
A more common approach is to train across multiple species (\tref{tab:existing_gLMs}), as typically done for pLMs.
As species become more distant, regulatory logic diverges faster than proteins.
One proposed approach is to explicitly add a species identifier as an extra input to the model \cite{speciesLM}.
Notwithstanding, it is plausible that a large enough model, with enough genomic context, might be able to naturally model distant genomes, similarly to how LLMs handle multilingual datasets.

As mentioned earlier, in prokaryotes, there exist models (MegaDNA and EVO) that take an entire genome as context \cite{megadna,evo}.
This is currently infeasible for eukaryotes, and therefore leads to the question of how to partition the genome into context windows to be separately modeled. 
Many interactions are restricted to nearby positions, such as transcription factor binding site motifs, motivating the development of models with a relatively small context ($<$ 6~kb) (\tref{tab:existing_gLMs}).
However, there are obvious long-range interactions, such as between exons of the same gene or between enhancers and promoters (up to 1~Mb) \cite{schoenfelder2019long}.
Such long context lengths introduce computational and statistical challenges, and efforts have been made to overcome them \cite{hyenadna,megadna,evo,caduceus}.
Regardless of the chosen context length, it is still not easy to partition the genome into independent units (similarly to how proteomes are separated by protein).
For instance, the enhancer of a gene can be located inside the intron of another gene \cite{schoenfelder2019long}, and multiple genes can be controlled by the same enhancer \cite{karnuta2018enhancers}.
Avoiding potential data leakage due to orthology and paralogy is quite challenging, especially when training across species.

The choice of training data may significantly influence gLMs' outputs and learned representations. DNA sequences observed in nature are the outcome of various evolutionary processes, the foremost of which are mutation and selection \cite{King1969}. For certain applications, it may be desirable to curate training data such that one of these processes is more manifest than the other. For example, for the sake of fitness prediction, it may be desirable to exclude/downweight hypermutable sites (such as CpG sites) and nonfunctional regions (such as certain classes of repetitive elements). 

\subsection*{Model architecture}
\label{sec:architecture}

CNN models \cite{deepbind, deepsea, basset, Basenji} have been widely used in genomics for supervised tasks prior to the emergence of the Transformer architecture \cite{transformer}. CNNs are particularly effective at capturing local dependencies and motifs within genomic sequences through their ability to apply filters across the input data. These models have been successful in predicting DNA-protein binding sites, regulatory elements, and TFBS. GPN \cite{gpn}, the aforementioned gLM for genome-wide variant effect prediction in \textit{Arabidopsis thaliana}, took inspiration from the success of language models with modified CNN layers in NLP \cite{tay2021pre} and protein modeling \cite{yang2024convolutions}, and replaced self-attention layers in a Transformer encoder with dilated CNN layers.

Transformer models have revolutionized various machine learning domains, particularly in NLP~\cite{transformer}, and have recently been widely adopted for genomics modeling.
The self-attention mechanism allows each token to attend to all positions in the input sequence simultaneously, enabling the model to dynamically focus on relevant parts of the sequence.
This capability has led to significant advancements in detecting regulatory mechanisms for supervised gene expression tasks \cite{enformer, borzoi}. 

Despite their strengths, Transformer models face several challenges unique to genomic modeling. One significant issue is that Transformers have weak or no inductive biases regarding the locality of interactions \cite{su2024roformer, bigbird}, making them less data-efficient at modeling local motifs such as TFBS. This motivated the development of CNN-Transformer hybrids such as LOGO \cite{LOGO}, following supervised models such as Enformer \cite{enformer}.

Another challenge is the context length: the self-attention mechanism results in computational time and memory scaling quadratically with the input sequence length, making it impractical to apply Transformers to very long genomic sequences \cite{dai2019transformer}. Consequently, the longest input length that conventional attention-based gLMs can handle so far is 12~kb for NT-v2 \cite{dalla2023nucleotide}. To address this limitation, several Transformer-based gLMs have implemented approximate attention or hierarchical attention methods that sacrifice full pairwise attention between all tokens. These methods include the use of sparse attention \cite{bigbird} in GENA-LM \cite{GENA-LM}, which extends the context length to 36~kb, and the MEGABYTE sub-quadratic hierarchical self-attention \cite{yu2024megabyte} employed in MegaDNA~\cite{megadna}, achieving a context length of 96~kb.

To overcome the quadratic scaling issues of self-attention, various state-space models (SSMs)~\cite{gu2021efficiently, poli2023hyena, gu2023mamba} have been proposed for gLMs as efficient alternatives to Transformers, offering nearly linear scaling with sequence length. HyenaDNA \cite{hyenadna}, based on the Hyena Hierarchy \cite{poli2023hyena}, can support input contexts as long as 1 million nucleotides. EVO \cite{evo}, a hybrid model combining Hyena and Transformer architectures, is pretrained with 8~kb sequences and later fine-tuned with 131~kb sequences during the context extension stage. Caduceus \cite{caduceus}, built on the Mamba-based  SSM \cite{gu2023mamba}, is trained on 131~kb sequences while 
incorporating reverse-complement equivariance.

\subsection*{Learning objective}
\label{sec:learning}

As described in Box~\ref{box:LM_framework}, the MLM task (sometimes also called ``masked token prediction'') involves predicting the identities of tokens randomly omitted from sequences with a predetermined probability (a common choice is 15\%) given the remaining tokens. This framework has been used to train the seminal LLM BERT \cite{devlin2018bert} and pLM ESM-1b \cite{rives2021biological}, and has since been widely used for training gLMs.
The CLM task (also referred to as ``autoregressive language modeling'' or ``next token prediction'') involves predicting the identities of tokens in sequences given their preceding tokens; it has been used to train the GPT series of LLMs \cite{brown2020language}. In this task, the model predicts the next token given the previous tokens in a unidirectional, left-to-right order.
A commonality between these two tasks is that they require models to predict components of data given other components as context. To generalize on these tasks, models must learn low-dimensional representations of the data. This capability enables the gLMs to understand and generate genomic sequences by capturing the underlying patterns and dependencies within the genome.
In protein modeling, MLM tends to achieve better representations and transfer learning capabilities than CLM \cite{cheng2024training}.
On the other hand, CLMs are the traditional choice for generation tasks, but excellent results have been recently obtained with MLMs via progressive unmasking \cite{samuel2024berts, esm-3}.

To reduce input sequence length and model longer context, both $k$-mer and byte-pair encoding~\cite{BPE} (BPE) \textit{tokenizations} create artificially defined nucleotide vocabularies larger than the natural nucleotide vocabularies of \texttt{\{A, C, G, T\}}.
On the other hand, single-nucleotide tokenization simplifies model interpretation and attribution, and enhances the model's ability to handle genomic variations more effectively.

Several modifications to the training objective have been explored to provide additional signal and boost performance. For instance, 
GPN-MSA \cite{gpnmsa} enhances MLM training on the human reference genome with a whole-genome MSA \cite{multiz,cactus} of  vertebrate species, leveraging conservation across related species for additional context. 
A limitation is that whole-genome MSAs have only been generated for certain species, and might require further development to be effective in plants~\cite{song2023new}.
Additionally, \textit{cis} regulatory elements tend to diverge fast in sequence space even if they have conserved activity, which limits the orthology information that can be extracted via alignment \cite{phan2024conservation}.
Species~LM \cite{speciesLM} directly integrates species information by assigning a dedicated token for each yeast species and appending the species token to the input sequence during training and inference.
Pretraining on nucleotide sequences has been expanded to enable cross-talk with additional modalities such as epigenetics \cite{GeneBERT,trotter2021epigenomic}, RNA \cite{zhu2024cd, he2024lucaone}, proteins \cite{zhu2024cd, cornman2024omg, he2024lucaone}, and natural language \cite{chatNT}.

\subsection*{Interpretation}
\label{sec:Interpretation}
Deep learning models, while having achieved remarkable performance in various prediction tasks, typically lack interpretability and are often used as ``black boxes''. However, understanding how these models generate such predictions is crucial for enabling broader applications and advancing model development. As a result, a series of methods have been developed to interpret deep learning models, including those specific to genomics \cite{linardatos2020explainable, zhang2021survey, talukder2021interpretation}. While the interpretation of gLMs is still an emerging line of research, several models have been shown to have learned meaningful biological patterns.

The sequence embeddings extracted from language models are commonly used as representations that capture rich contextual information and sequence features. Unsupervised clustering of the encoded sequence embeddings from gLMs has shown distinct clusters of input sequences that correspond to different genomic classes such as CDS, intronic, UTR, etc.~\cite{dnabert, gpn, dalla2023nucleotide, hyenadna} (\fref{fig:applications}d). Additionally, unsupervised clustering of SpliceBERT embeddings of canonical splice sites and non-splice GT/AG sites reveals distinct clusters that correspond to the two groups  \cite{SpliceBERT}. 
These results suggest that the models have learned to capture key contextual patterns that characterize functional elements in the genome. 

The attention mechanism in the Transformer model is designed to capture the pattern of interaction between input tokens. Thus, interpreting the attention weights or the attention map for a given input sequence can reveal genomic features learned by the model.  In SpliceBERT \cite{SpliceBERT},  attention weights between splice donors and acceptors are significantly higher than those between random pairs of sites; also, the strength of interaction tends to be higher within true donor-acceptor pairs compared to other combinations of donor and acceptor sites.  These findings suggest that the model has learned the relationship between functionally interacting sites.

The nucleotide reconstruction approach has also been used in several gLMs to discover sequence motifs learned by the models. Specifically, individual positions of the input sequence are masked one at a time and the probability distribution of the nucleotides is predicted by the trained model given the genomic context. The obtained distribution at each site can reveal motifs learned by the model.
This approach has been used in GPN to find notable patterns in the distribution of the reconstructed nucleotides.  In particular, the model’s predictions are generally more confident in functionally important sites. For example, coding sequences and splice donor/acceptor sites are typically predicted with higher confidence than deep intronic sites. Moreover, within coding sequences, the third nucleotide position of a codon, the least determinant of the translated amino acid, is typically predicted with lower confidence than the first two nucleotide positions. Adapting TF-MoDISco \cite{modisco}, a dedicated tool to identify novel TFBS using model predictions, the authors also found sequence motifs that match known ones in TFBS databases and relevant literature~\cite{gpn} (\fref{fig:applications}a).
Similarly, the reconstructed sequence motifs from Species~LM \cite{speciesLM} also match the binding sites of known DNA- and RNA-binding proteins in species that are unseen during training, with the fidelity of motif reconstruction depending on the context and genomic regions that correctly reflect the \textit{in vivo} binding sites.  Furthermore, the reconstructed motifs' composition, existence, and location exhibit species-specific patterns, which suggests gLM as a potentially powerful tool for investigating the evolution of sequence motifs and regulatory code.

More recently, the dependency between genomic positions learned by a gLM was studied by introducing point mutations at a position and quantifying the changes in nucleotide probabilities at other positions \cite{tomaz2024nucleotide}.
Nucleotide dependency analysis revealed learned interactions within and across functional elements such as TFBS, splice sites and RNA, including known secondary and tertiary structure contacts.
Notably, nucleotide dependency analysis was able to detect bound TFBS more robustly than the previous approach based on predicted marginal probability distributions.

\subsection*{Evaluation}
\label{sec:evaluation}
In this section, we discuss how models' performance can be benchmarked in regards to the three application areas described earlier:
predicting functional constraints on alleles, generating novel viable sequences, and transfer learning.

There are various types of data that reflect functional constraint on alleles and may be used to benchmark a variant effect predictor. One type of data are assays that couple functional differences between genetic variants to readouts (such as the expression of a reporter gene or cell growth) \cite{fowler2023atlas, Kircher2019Saturation, findlay2018brca1sge}. These readouts may be used to rank variants by their functionality, and since variants that affect function also tend to be under selection, we should expect that these ranks should correlate with ranks obtained from models' predictions. One source for these data is ProteinGym, a widely-used collection of experimental data that may be used to benchmark missense variant effect predictors \cite{Notin2023ProteinGym}. Another type of data are clinical labels indicating whether variants have evidence of pathogenicity---that is, can elevate the risk for diseases. Pathogenic variants may affect fecundity, and, therefore, be deleterious. As a result, we can benchmark variant effect predictors by evaluating them as pathogenicity classifiers.  In human genetics, primary sources of clinical labels for variants include the ClinVar \cite{landrum2016clinvar}, HGMD \cite{Stenson2017}, and OMIM \cite{amberger2015omim} databases. A third type of data are variant frequencies. Since common variants are unlikely to be highly deleterious \cite{pritchard2002allelic}, their predicted level of constraint should be relatively higher than those of rare variants. Therefore, we may benchmark predictors based on how well they identify common variants. A primary source of data on human allele frequencies in various ancestry groups is the gnomAD database \cite{Karczewski2020MutationalConstraint}. Altogether, these data may be used as separate lines of evidence for models' generalization performance.

\begin{mybox}[t!]
\begin{mdframed}[style=box]
\caption{Evaluating Generalization Performance}
\label{box:generalization}
The purpose of evaluating predictive models is to build trust in their capability to generalize -- that is, to make satisfactory predictions for unlabeled data. A straightforward and standard way to estimate the generalization performance of a model is to evaluate its accuracy on a “test set” of labeled data that are representative of unlabeled data of interest \cite{vapnik1999nature}. This approach is the basis of most machine learning benchmarks.

\medskip
Importantly, for this evaluation to be a reliable indicator of generalization performance, models must not be provided any information that may be used to differentiate test set data from the data they will ultimately be deployed on. Otherwise, they may decrease their test set error at the expense of their generalization performance. For this reason, machine learning contests that withhold their test data from participants are routinely organized \cite{ILSVRC15,moult2018critical,johnson2017cagi}.
\end{mdframed}
\end{mybox}

An issue with the evaluation of variant effect predictors is that the relationship between validation data and functional constraint may be murky. As a consequence, models may excel at benchmarks by exploiting the ways in which data fail to capture functional constraint. For example, a critical issue with using clinical labels is that variants are classified based on whether there is ample evidence that they are benign or pathogenic \cite{grimm2015evaluation}. Since predictors may also utilize this evidence, their benchmarked performance on labeled variants may not reflect their true performance on unlabeled variants. 
(See Box~\ref{box:generalization} for a brief discussion of generalization performance.)
There are also critical issues with using allele frequency data: for one, in addition to the direct action of natural selection, allele frequencies are influenced by factors such as mutation rates, drift, background selection, and hitchhiking \cite{hartl1997principles}.
As a result, predictors may perform well on benchmarks by predicting the effects of these processes instead of functional constraint. These issues highlight a need to carefully interpret the causes of predictors' performance, and they have led to calls for greater transparency on which data and methods are used to train predictors \cite{livesey2024guidelines}.

There are a separate set of challenges with the evaluation of generative sequence models. A basic way to evaluate language models' generative capabilities is to compare their perplexities on sets of valid sequences. However, to evaluate models' capability to design novel sequences, it is necessary to gauge whether they can identify sequences that are both viable and novel. For this reason, models' perplexities on test sets may not reliably indicate their utility for design. Instead, a holistic approach that examines a broad range of properties of generated sequences may be warranted. For instance, Polygraph \cite{gupta2023polygraph}, a recent benchmark for regulatory sequence design, proposes a series of analyses that investigate sequence composition, motif patterns, and predicted functional activity. For whole-genome or chromosome design tasks, it may also be necessary to evaluate the existence and positioning of essential genes and functional regulatory elements, as well as the interactions between them. Ultimately, the designed sequences should be experimentally evaluated to determine if they perform their desired functions.

Lastly, there is a unique challenge with the evaluation of gLMs for transfer learning: any set of benchmarks---perhaps, in conjunction---must reliably indicate a model's performance on relevant tasks. A type of data that may be fashioned into a set of tasks broadly informative of models' adaptability to genome interpretation are functional genomics data (such as those from the ENCODE \cite{encode2012integrated} or Roadmap Epigenomics \cite{kundaje2015integrative} projects), which may be used to annotate genomic regions and variants. We should expect that a model's performance on predicting these annotations from genome sequences after adaptation is indicative of their capability to identify functionally similar genomic elements. To facilitate comparison between models, these annotations have been consolidated into various standardized sets of training and test data \cite{dalla2023nucleotide, Gresova2023, robson2023guanine, marin2024bend}. 

As transfer learning benchmarks help highlight limitations of current models and establish criteria for publication, they are likely to be important assets for gLM developers and users. However, despite differences in current benchmarks' choice of tasks and methodologies, they provide seemingly redundant insight into gLMs' capabilities. Moving forward, it will be incumbent on the computational genomics community to develop standardized and extensible benchmarks that are widely trusted.

\section*{CONCLUDING REMARKS \& FUTURE PERSPECTIVES}
In an age of a vast and growing number of genomic sequences, gLMs are emerging as powerful tools to extract complex patterns useful for numerous applications, including functional constraint estimation, sequence design and transfer learning. 
However, they do not yet represent a magical, sudden breakthrough as the term ``AI'' may suggest.  Instead, we view them as another useful modeling tool, much like Hidden Markov Models were when they were first introduced.
Often, gLMs are claimed to be ``foundation models'', a term recently invented to denote models trained on broad data that can be adapted to a wide range of downstream tasks \cite{bommasani2021a}.
The introduction of this new term has been criticized as the word ``foundation'' has the connotation of substantial improvement on downstream task performance, which is an empirical question, not an inherent property of pretrained models \cite{helfrich2024harms}.
This criticism rings even louder in new domains such as genomics, where establishing adequate benchmarks is likely to take some time.

While earlier gLMs tend to be more or less direct adaptations from NLP models, we expect that further contextualization with deep genomics expertise will reap the highest rewards.  
We note that evaluating the capabilities of gLMs is challenging because metrics may be misleading, especially when over-optimized. A boon for NLP is that humans are experts in natural language and, therefore, can calibrate benchmarks to match their expertise. In genomics, however, we must rely on data and expert knowledge to falsify models. This aspect of the problem makes it especially challenging and may highlight a need for engagement with subject-matter experts and deliberate experimentation for the sake of developing benchmarks.  
We conclude this review with a few research directions (listed in Outstanding Questions)
that we believe warrant further investigation.

\begin{mdframed}[style=box]
{\large Outstanding Questions}
\begin{enumerate}[leftmargin=5mm]
\item
How can we best model patterns across a wide range of scales, from motifs to genes to whole genomes?
\item
For which applications is it important to model long-range interactions and how does one determine a suitable size of the receptive field?
\item
How can we incorporate structural variations into gLMs?
\item
What is the best way to utilize population genetic data when training gLMs?
\item
How can we best integrate gLMs with other complex modalities, such as transcriptomic and epigenetic data?
\item
For developing gLMs, can we better understand what makes some genomes harder to model than others?
\item 
Will the scaling hypothesis hold for gLMs, and for how long?
Are there really that much data available, considering that most may be non-functional?
\end{enumerate}
\end{mdframed}

\subsection*{Acknowledgments}
This work is supported in part by an NIH grant R35-GM134922, a grant from the Koret-UC Berkeley-Tel Aviv University Initiative in Computational Biology and Bioinformatics, and a grant from the Noyce Initiative UC Partnerships in Computational Transformation Program.

\subsection*{Declaration of Interests}
The authors declare no competing interests.

\clearpage
\bibliographystyle{numbered}
\bibliography{references}

\end{document}